\documentclass[aps,eqsecnum]{revtex4}
\usepackage{epsfig}
\usepackage{graphicx}
\usepackage{amsmath}
\usepackage{amssymb}
\usepackage{color}
\newif\ifold             \oldtrue            
\def\ba{\begin{eqnarray}}
\def\ea{\end{eqnarray}}

\newcommand{\be}{\begin{equation}}
\newcommand{\ee}{\end{equation}}

\begin{document}
\title{Gap generation and semimetal-insulator phase transition in  graphene}
\author{O. V. Gamayun,\, E. V. Gorbar,\, and V. P. Gusynin}
\email{gamayun@bitp.kiev.ua, gorbar@bitp.kiev.ua, vgusynin@bitp.kiev.ua}
\affiliation{Bogolyubov Institute for Theoretical Physics, 14-b Metrologichna
str., Kiev 03680, Ukraine}

\begin{abstract}

The gap generation is studied in suspended clean graphene in the
continuum model for quasiparticles with the Coulomb interaction. We solve the
gap equation with the dynamical polarization function and show that, comparing
to the case of the static polarization function,  the critical
coupling constant  lowers to the value $\alpha_{c}=0.92$, which is close to that obtained in
lattice Monte Carlo simulations. It is argued that additional short-range
four-fermion interactions should be included in the continuum model
to account for the lattice simulation results. We obtain the critical
line in the plane of electromagnetic and four-fermion coupling constants and
find a second order phase transition  separating
zero gap and gapped phases with critical exponents close to those found in
lattice calculations.

\end{abstract}

\maketitle

\section{Introduction}

 Graphene, a one-atom-thick layer of graphite, is a remarkable system with many unusual
properties that was fabricated for the first time  five years ago \cite{graphene-fabrication}.
Theoretically it was shown
long time ago \cite{gr} that quasiparticle excitations in graphene have a linear
dispersion at low energies and are described by the massless Dirac equation in 2+1 dimensions.
The observation of anomalous integer quantum Hall effect in graphene
\cite{QHE-experiment} is in perfect agreement with the theoretical
predictions \cite{QHE-theory} and became a direct experimental proof of the existence of
gapless Dirac quasiparticles in graphene.

The  unusual band structure of graphene
 has an important consequence for the electron-electron interaction in this material.
In the continuum limit, graphene model on a honeycomb lattice, with both on-site and
nearest-neighbor repulsions, maps onto a (2 + 1)-dimensional field theory of Dirac fermions
interacting through the Coulomb potential plus, in general, some residual short-range interactions
represented by local four-fermion terms. The vanishing density
of states at the Dirac points ensures that the Coulomb interaction between the
electrons in graphene retains its long-range character due to vanishing of the
static polarization function for $q \to 0$ \cite{Gonzalez}. The large value of the
``fine structure'' coupling constant $\alpha=e^2/\hbar v_F \sim 1$ means that a strong attraction
takes place between electrons and holes in graphene at the Dirac points.
As is known, for graphene on a substrate with the effective coupling $\alpha/\kappa\ll1$, $\kappa$
being a dielectric constant,  the system is in a weak coupling regime and exhibits
semimetallic properties due to the absence of a gap in the electronic spectrum.
Much less is known about  suspended graphene where the coupling constant is large.
In fact, suspended graphene provides a condensed-matter analogue of strongly coupled
quantum electrodynamics (QED) intensively studied in the 70-ties and 80-ties
\cite{Fomin,review,Bardeen,Kogut,Miranskybook}.
The  dynamics of the vacuum in QED leads to many peculiar effects not yet observed
in nature. Some QED-like effects such as zitterbewegung (trembling motion) \cite{Katsnelson},
Klein tunneling \cite{Klein}, Schwinger pair production \cite{Schwinger-effect},
supercritical atomic collapse \cite{Shytov1,Fogler},
 have a chance to be tested in graphene (for experimental
observation of the Klein tunneling in graphene, see \cite{Kim}).
To observe these effects in graphene, it is important to use
suspended and clean samples where charges from a substrate do not interfere
with the dynamics of electrons.

Recently, it is was shown in Ref.\cite{excitonic-instability} that,
for strong enough coupling $\alpha > \alpha_c$, there is a tachyonic solution
in the spectrum of the Bethe--Salpeter (BS) equation for the electron-hole bound
state signaling the presence of excitonic instability of the zero-gap ground
state of monolayer graphene in the supercritical regime.
The critical coupling equals $\alpha_{c}=1/2$ if the vacuum polarization is neglected
and $\alpha_{c}\approx1.62$ in the random phase approximation with the static
polarization  \cite{footnote1}.
It was also shown there  that physically the
excitonic instability is connected with the well-known supercritical Coulomb
center problem \cite{Popov} where $\alpha_{c}=1/2$ in two spatial dimensions
\cite{footnote3}. The situation is similar to that in the
theory of superconductivity \cite{Schrieffer}, where the four-fermion vertex
instability has its origin in the Cooper pair problem. It was argued in
\cite{excitonic-instability} that the formation of an excitonic condensate of
electron-hole pairs should cure the excitonic instability and lead to opening
of a quasiparticle gap in a free standing clean graphene resulting in
dramatic changes in the transport properties. A similar situation
occurs in QED in 3+1 dimensions where the gap generation takes place in
the strong coupling regime \cite{Fomin,review,Miranskybook} (see, also,
\cite{Bardeen,Kogut}).

The problem of gap generation in graphene was considered
before the actual fabrication of this material in Refs.
\cite{Khveshchenko,Gra2002,GGMS} where the random phase approximation with
the static polarization function was used. Recently,
lattice Monte Carlo simulations  found the value of
the critical coupling $\alpha_{c}=1.08$ for a semimetal-insulator
transition \cite{Drut} and this motivated
us to reconsider the problem of gap generation in graphene.  It was
already indicated in \cite{Gra2002} that taking into account the frequency
dependent polarization function should lower the critical coupling. We
investigate this question in the present paper and confirm that the dynamical
polarization is indeed quantitatively important :
solving the gap equation with frequency dependent polarization function
we find the critical coupling $\alpha_c=0.92$ instead of $\alpha_{c}=1.62$
in the case of static polarization. We would like to note also that the
presence of a gap would be valuable for electronics
applications, in particular, for working graphene transistors.

Another problem studied in the present paper is the order of phase
transition connected with the gap generation in graphene. Due to the scale
invariance of the model with the Coulomb interaction an infinite order phase
transition was found in \cite{Gra2002,GGMS}. Such a phase transition belongs to
the class of the so-called conformal phase transitions \cite{CPT}. According to
the recent Monte Carlo (MC) simulations \cite{Drut} (for related MC simulations,
see Ref.\cite{Hands}), the semimetal-insulator phase
transition in graphene is of the second order. One of the reasons for such
a difference might be lattice finite size effects which can change the
order of phase transition \cite{GusyninReenders,Fisher,Liu}.
On the other hand, according to \cite{Alicea,Herbut,Aleiner}, the effective
continuum theory for quasiparticles in graphene should contain besides the Coulomb
interaction some additional contact four-fermion interaction terms that arise
from the microscopic graphene lattice interactions. These terms contain a dimensionful
parameter, therefore, they explicitly break the scale invariance
of the continuum model. In such a case, one may expect a conventional second
order phase transition. In order to take into account these four-fermion
interaction terms, we consider in the present paper the simplest Gross--Neveu
interaction term and show that the presence of this interaction term
 plays an important role: First, instead of a critical point we now have a
critical line in the plane of electromagnetic and four-fermion coupling constants
separating symmetric and symmetry broken phases.  Second, the inclusion of this
term indeed changes the order of phase transition from infinite to the second
order along a part of the critical line $0<\alpha<\alpha_{c}$.
Third, it lowers the value of the critical electromagnetic
coupling comparing to the case of purely Coulomb interaction.
At last, the critical indices stay closer to those
obtained in lattice simulations \cite{Drut}.

The structure of the paper is the following. We begin with presentation in Sec. \ref{model}
of the continuum model describing graphene quasiparticles interacting through the Coulomb
potential. In Sec. \ref{gap-equation} we solve the gap equation with the frequency
dependent one-loop polarization function and determine a critical coupling for the onset
of a gap. To get insight into analytical solutions of the gap equation,  we then turn back
to the case of the static polarization and find asymptotical behavior of the gap function,
calculate the excitonic condensate $\langle\overline\Psi\Psi\rangle$ of
particle-hole pairs, the correlation length, and critical exponents near
the phase transition point $\alpha_{c}$. In Sec. \ref{phase-diagram}
we include the Gross--Neveu four-fermion interaction, find
explicitly the critical line in the plane of Coulomb and four-fermion
interaction coupling constants, and determine the critical exponents for the
phase transition along this line. In Conclusion we summarize the
main results.

\section{The model}
\label{model}

For the description of the dynamics in graphene, we will use the
same model as in Refs.\cite{Khveshchenko,Gra2002}  in which while quasiparticles are confined to
a two-dimensional plane,  the electromagnetic (Coulomb) interaction between
them is three-dimensional in nature. The low-energy quasiparticles excitations in
graphene are conveniently described in terms of a four-component Dirac spinor
$\Psi^{T}_{a}=(\psi_{KAa},\psi_{KBa},\psi_{K^{\prime}Ba},\psi_{K^{\prime}Aa})$
which combines the Bloch states with spin indices $a=1,2$
on the two different sublattices $(A, B)$ of the hexagonal graphene lattice and
with momenta near the two-nonequivalent valley points $(K, K^{\prime})$ of the
two-dimensional Brillouin zone. In what follows we treat the spin index as a ``flavor''
index with $N_{f}$ components, $a=1,2,\dots N_{f}$, then $N_{f}=2$ corresponds
to graphene monolayer while $N_{f}=4$ is related to the case of two
decoupled graphene layers, interacting solely via the Coulomb
interaction.

The action describing graphene quasiparticles interacting
through the Coulomb potential has the form
\ba
S&=&\int dtd^{2}r\overline{\Psi}_{a}(t,\mathbf{r})\left(i\gamma^{0}\partial_{t}-iv_{F}
{\mathbf{\gamma}}{\mathbf{\nabla}}\right)\Psi_{a}(t,{\mathbf{r}})\nonumber\\
&-&\frac{1}{2}\int dt dt^{\prime}
d^{2}rd^{2}r^{\prime}\overline\Psi_{a}(t,\mathbf{r})\gamma^{0}\Psi_{a}(t,\mathbf{r})
U_{0}(t-t^{\prime},|\mathbf{r}-\mathbf{r}^{\prime}|)
\overline\Psi_{b}(t^{\prime},\mathbf{r}^{\prime})\gamma^{0}\Psi_{b}(t^{\prime},\mathbf{r}^{\prime}),
\label{action}
\ea
where $v_{F}$ is the Fermi velocity, $\overline\Psi=\Psi^{\dagger}\gamma^{0}$,
and the $4\times4$ Dirac $\gamma$-matrices
$\gamma^{\mu}=\tau^3\bigotimes \, (\sigma^3,i\sigma^2,-i\sigma^1)$ furnish a reducible
representation of the Dirac algebra in $2+1$ dimensions. The bare Coulomb potential
$U_{0}(t,|\mathbf{r}|)$ is given by
\be
U_{0}(t,|\mathbf{r}|)=\frac{e^{2}\delta(t)}{\kappa}\int \frac{d^{2}k}{2\pi}\frac{e^{i\mathbf{k}\mathbf{r}}}
{|\mathbf{k}|}=\frac{e^{2}\delta(t)}{\kappa|\mathbf{r}|}.
\ee
However, the polarization effects considerably modify this bare Coulomb potential and the interaction
will be
\be
U(t,|\mathbf{r}|)=\frac{e^{2}}{\kappa}\int\frac{d\omega}{2\pi}\int \frac{d^{2}k}{2\pi}
\frac{\exp(-i\omega t +i\mathbf{k}\mathbf{r})}{|\mathbf{k}|+\Pi(\omega,\mathbf{k})},
\ee
where $\kappa$ is the dielectric constant due to a substrate and the polarization function $\Pi(\omega,\mathbf{k})$
is proportional (within the factor $2\pi/\kappa$) to the time component of the photon polarization function.
Correspondingly,  the Coulomb propagator has the form
\be
D(\omega,\mathbf{q})=\frac{1}{|\mathbf{q}|+\Pi(\omega,\mathbf{q})},
\label{Coulombpropagator}
\ee
where the one-loop polarization function  is \cite{Gonzalez}
\be
\Pi(\omega,\mathbf{k})=\frac{\pi e^{2}N_{f}}{4\kappa}\frac{\mathbf{k}^{2}}{\sqrt{\hbar^{2}
v_{F}^{2}\mathbf{k}^{2}-\omega^{2}}},
\label{1loop-polarization}
\ee
and in the instantaneous approximation it becomes
\be
\Pi(\omega=0,\mathbf{k})=\frac{\pi e^{2}N_{f}}{4\kappa\hbar v_{F}}|\mathbf{k}|.
\label{instant-approx}
\ee
The continuum effective theory described by the action (\ref{action}) possesses
$U(2N_{f})$ symmetry. In the case of graphene, $N_{f}=2$, the corresponding
16 generators are (see, for example, Ref.\cite{Gra2002}):
\be
\frac{\sigma^{\alpha}}{2}\otimes I_{4},\quad \frac{\sigma^{\alpha}}{2i}\otimes\gamma^{3},
\quad \frac{\sigma^{\alpha}}{2}\otimes\gamma^{5}, \quad \frac{\sigma^{\alpha}}{2}
\otimes\gamma^{3}\gamma^{5},
\label{symmetry}
\ee
where $I_{4}$ is the $4\times4$ Dirac unit matrix, and $\sigma^{\alpha}$, with $\alpha=0,1,2,3$,
are four Pauli matrices connected with the spin degrees of freedom ($\sigma_{0}$ is
the $2\times2$ unit matrix). However, as was pointed out in Ref. \cite{Alicea} (see
also Refs. \cite{Herbut,Aleiner}), this symmetry is not exact in the graphene
tight-binding model on lattice. In fact, there are small on-site  interaction terms
which break the $U(2N_{f})$ - symmetry, their role will be considered in Sec.
\ref{phase-diagram}.

\section{Gap generation and the critical coupling constant}
\label{gap-equation}

In this section we study spontaneous generation of a gap in the
quasiparticle spectrum of graphene. The Schwinger-Dyson equation for the quasiparticle
propagator has the form,
\be
S^{-1}(p_{0},\mathbf{p})=p_{0}\gamma^{0}-\mathbf{p}\pmb{\gamma}-ie^{2}\int\frac{d^{3}k}{(2\pi)^{2}}
D(p_{0}-k_{0},\mathbf{p}-\mathbf{k})\gamma^{0}S(k_{0},\mathbf{k})\gamma^{0},
\ee
where the Coulomb propagator $D(q_{0},\mathbf{q})$ is given by Eq.(\ref{Coulombpropagator})
and in the random phase approximation the polarization is taken as in
Eq.(\ref{1loop-polarization}). The vertex corrections are rather small \cite{Gonzalez}
and we neglect them in what follows.

\begin{figure}[ht]
\includegraphics[width=6.5cm]{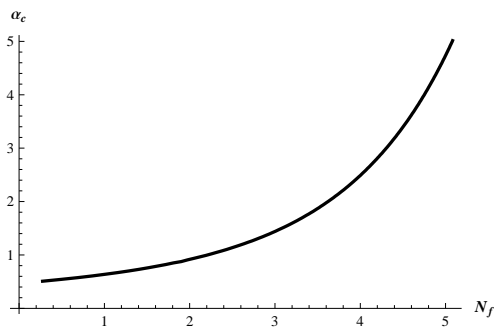}
\caption{The critical coupling as a function of $N_{f}$.}
\label{critline}
\end{figure}

The general form of the propagator of quasiparticles is
\be
S^{-1}(p_{0},\mathbf{p})=Z^{-1}p_{0}\gamma^{0}-A\mathbf{p}\pmb{\gamma} - \Delta,
\ee
where $Z,A,\Delta$ are functions of $p_{0},\mathbf{p}$ and we included also a bare gap
$\Delta_{0}$. We assume that a dependence of
these functions on the energy $p_{0}$ is rather weak so that we can approximate these functions
by their values at $p_{0}=0$. In this approximation it is easy to see that $Z=1$,
then after the Wick rotation, $k_{0}=i\omega$, we get a coupled system of equations
for $A(\mathbf{p}),\Delta(\mathbf{p})$:
\ba
A(\mathbf{p})&=&1+\frac{e^{2}}{\kappa\mathbf{p}^{2}}\int\limits_{-\infty}^{\infty}d\omega
\int\frac{d^{2}k}{(2\pi)^{2}}D(\omega,
\mathbf{p}-\mathbf{k} )\frac{\mathbf{p}\mathbf{k}A(\mathbf{k})}{\omega^{2}+\mathbf{k}^{2}
A^{2}(\mathbf{k})+\Delta^{2}(\mathbf{k})},
\label{eq:A}\\
\Delta(\mathbf{p})&=&\Delta_{0}+\frac{e^{2}}{\kappa}\int\limits_{-\infty}^{\infty}d\omega\int
\frac{d^{2}k}{(2\pi)^{2}}D(\omega,\mathbf{p}-\mathbf{k} )\frac{\Delta(\mathbf{k})}
{\omega^{2}+\mathbf{k}^{2}A^{2}(\mathbf{k})+\Delta^{2}(\mathbf{k})}.
\label{eq:Delta}
\ea
We write the integral over $\omega$ as
\be
I=\int\limits_{-\infty}^{\infty}d\omega D(\omega,\mathbf{q})
\frac{1}{\omega^{2}+\mathbf{k}^{2}A^{2}
+\Delta^{2}}=\int\limits_{-\infty}^{\infty}\frac{dx\,f(x)}{x^{2}\mathbf{q}^{2}+\mathbf{k}^{2}A^{2}
+\Delta^{2}},\quad f(x)=\frac{\sqrt{x^{2}+1}}{\sqrt{x^{2}+1}+g},\quad g=\pi N_{f}\alpha/4.
\label{int-over-omega}
\ee
The function $f(x)$ changes slowly from $1/(1+g)$ at $x=0$ (the instantaneous
approximation for $D(\omega,\mathbf{q})$) up to $1$ at $x=\infty$. The integral in
Eq.(\ref{int-over-omega}) can be evaluated exactly,
\ba
I=\frac{1}{|\mathbf{q}|\sqrt{\mathbf{k}^{2}A^{2}+\Delta^{2}}}J(d,g),\quad
d=\frac{\sqrt{\mathbf{k}^{2}A^{2}+\Delta^{2}}}{|\mathbf{q}|},\quad
J(d,g)=\frac{(d^{2}-1)(\pi-g c(d))+d g^{2}c(g)}{d^{2}+g^{2}-1},
\label{Jfunction}
\ea
where
\be
c(x)=\frac{2\cosh^{-1}(x)}{\sqrt{x^{2}-1}},\quad x>1,\quad c(x)=\frac{2\cos^{-1}(x)}{\sqrt{1-x^{2}}},
\quad x<1,\quad c(1)=2.
\ee
For $\Delta=0$ and setting $A=1$ on the right hand side of Eq.(\ref{eq:A}) we get
the leading one-loop correction \cite{velocity-renorm}, which comes from the range
of momenta $k\gg p$ in the integral,
\be
A(p)=1+\frac{2}{\pi^{2}gN_{f}}\left(\pi-2 g+(g^{2}-1)c(g)\right)\ln\frac{\Lambda}{p}+\mbox{finite terms},
\ee
where $\Lambda$ is a momentum cutoff of order the inverse lattice spacing in graphene.
The function $A(p)$ renormalizes the Fermi velocity $v^{*}_{F}(p)=v_{F}A(p)$.
The growth of $v^{*}_{F}(p)$ in the infrared stops when a non-zero quasiparticle gap
is taken into account (see, Eq.(\ref{eq:A})). In what follows we assume that the velocity
renormalization is already performed \cite{footnote4}
and put $A=1$ in Eq.(\ref{eq:Delta}) which then takes the form
\be
\Delta(\mathbf{p})=\Delta_{0}+\frac{e^{2}}{\kappa}\int\frac{d^{2}k}{(2\pi)^{2}}\frac{\Delta(\mathbf{k})}
{|\mathbf{p}-\mathbf{k}|\sqrt{\mathbf{k}^{2}+\Delta^{2}(\mathbf{k})}}J(d=\frac{|\mathbf{k}|}
{|\mathbf{p}-\mathbf{k}|},g),
\label{eq1:Delta}
\ee
where we set also $\Delta=0$ in the variable $d$. Since
the function $J$ depends weakly on the angle between the vectors $\mathbf{p}$ and $\mathbf{k}$,
 we can approximate $|\mathbf{p}-\mathbf{k}|\rightarrow {\rm max}(|\mathbf{p}|,
|\mathbf{k}|)$. Thus we  write
\be
J\left(\frac{k}{{\rm max}(k,p)},g\right)=J(1,g)\theta(k-p)+J\left(\frac{k}{p},g\right)\theta(p-k).
\ee
Assuming $\Delta(\mathbf{p})=\Delta(|\mathbf{p}|)$ and
integrating over the angle in Eq.(\ref{eq1:Delta}), we get
\ba
\Delta(p)=\Delta_{0}+\frac{\alpha}{\pi^{2}}\int\limits_{0}^{\Lambda}\frac{dk\, k\,\Delta(k)}{\sqrt{k^{2}
+\Delta^{2}(k)}}{\cal K}(p,k),
\label{gap:eq}
\ea
where the kernel
\ba
{\cal K}(p,k)=\frac{\theta(p-k)}{p}K\left(\frac{k}{p}\right)J\left(\frac{k}{p},g\right)
+\frac{\theta(k-p)}{k}K\left(\frac{p}{k}\right)J(1,g),
\label{kernel-identity}
\ea
and $K(x)$ is the complete elliptic integral of the first kind,  $\theta(x)$
is the Heaviside step function.
For zero bare gap, $\Delta_{0}=0$, Eq.(\ref{gap:eq})  admits a nontrivial solution
which bifurcates from the trivial one at some critical coupling $\alpha=\alpha_{c}$.
To find this critical point we neglect the terms quadratic or higher order in $\Delta$
in Eq.(\ref{gap:eq}). It must be emphasized that this is {\em not} an approximation:
it is a precise manner to locate the critical point by applying bifurcation theory \cite{bifurcation}.
Hence the bifurcation equation amounts to a linearization of Eq.(\ref{gap:eq})
with respect to the gap function, the result reads
\be
\Delta(p)=\frac{\alpha}{\pi^{2}}\int\limits_{0}^{\infty}d k\,\Delta(k)
\,{\cal K}(p,k).
\ee
Note that the ultraviolet cutoff, $\Lambda$, has been taken to infinity, which is appropriate
at the bifurcation point \cite{bifurcation}. This equation is scale invariant and  is solved by
$\Delta(p)=p^{-\gamma}$ on the condition that the exponent $\gamma$ satisfies
the transcendental equation
\ba
1&=&\frac{4g}{\pi^{3}N_{f}}\int\limits_{0}^{\infty}dx\,x^{-\gamma}\left[\theta(1-x)K(x)J(x,g)+
J(1,g)\frac{\theta(x-1)}{x}K\left(\frac{1}{x}\right)\right]\nonumber\\
&=&\frac{4g}{\pi^{3}N_{f}}\int\limits_{0}^{1}dx
\left[x^{-\gamma}J(x,g)+J(1,g)x^{\gamma-1}\right]K(x),\quad 0<\gamma<1,
\label{trans:eq}
\ea
where $J(x,g)$ is given by Eq.(\ref{Jfunction}), and
\be
J(1,g)=\left\{\begin{array}{c}\frac{2\arccos g}{\sqrt{1-g^{2}}},\quad g\le 1,\\
\frac{\ln(g+\sqrt{g^{2}-1})}{\sqrt{g^{2}-1}},\quad g\ge 1.\end{array}\right.
\ee
Eq.(\ref{trans:eq}) defines roots $\gamma$ for any value of the coupling $g$. A bifurcation
occurs when two roots in $(0,1)$ become equal. Numerically we find that this happens for
$\gamma=1/2$ and  the critical coupling ($N_{f}=2$),
\be
g_{c}=1.445,
\ee
which corresponds to $\alpha_{c}=0.92$.
For values $g>g_{c}$ the roots become complex indicating that oscillatory behavior
of the gap function takes over from non-oscillatory one. Equation (\ref{trans:eq})
determines the critical line in the plane $(\alpha,N_{f})$ which is presented in
Fig.\ref{critline}.
This line should be compared with the critical line
\be
\alpha_{c}=\frac{4\lambda_{c}}{2-\pi N_{f}\lambda_{c}}
\label{critline-Coulomb}
\ee
obtained in Ref.\cite{Gra2002} using the static polarization function ($\lambda_{c}
=1/4$ in Ref.\cite{Gra2002} for the kernel approximation (\ref{kernel-approximation})
used below, and $\lambda_{c}=0.23$ for more refined bifurcation analysis in Ref.
\cite{excitonic-instability}).
The most crucial difference between two
critical lines is that there is a critical number of flavors, $N_{crit}=2/\pi\lambda_{c}$,
for the critical line (\ref{critline-Coulomb}) for which
$\alpha=\infty$ while $\alpha$ never tends to infinity at finite $N_{f}$ for
 the critical line (\ref{trans:eq}) presented in Fig.\ref{critline}.

Recently, in  Ref.\cite{Khveshchenko1}  an approximation for the frequency dependent one-loop
polarization (\ref{1loop-polarization}) was used which reduces it to  (\ref{instant-approx})
 with additional $\sqrt{2}$ in the denominator, in this case the critical value $\alpha_{c}=1.13$.
The more refined analysis using bifurcation theory gives
$\alpha_{c}=0.93$ very close to the value we found above. We remind
also that renormalization group calculations in two loops yield $\alpha_{c}=0.833$
\cite{Vafek}.

A dynamical gap is generated only if $\alpha>\alpha_{c}$. Since for
suspended clean graphene the ``fine structure'' constant
$\alpha\approx 2.19$ is supercritical, the dynamical gap will be
generated making graphene an insulator. Note that for graphene on a
SiO$_{2}$ substrate the dielectric constant $\kappa\approx2.8$ and
$\alpha\approx0.78$, i.e., the system is in the subcritical regime. The value
of $\alpha_{c}$ is rather large that implies that a weak coupling
approach might be quantitatively inadequate for the problem of the gap
generation in suspended clean graphene. Therefore, it is instructive
to compare our analytical results with lattice Monte Carlo studies \cite{Drut},
where $\alpha_{c}=1.08\pm0.05$ for $N_{f}=2$ that is rather close to our
analytical findings.

\section{Nonlinear equation and critical exponents}
\label{NL-equation}

\begin{figure}[ht]
\includegraphics[width=8cm]{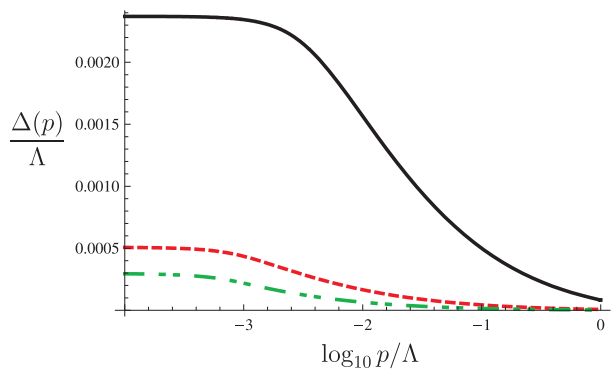}
\caption{Momentum dependence of the solution to the gap equation (\ref{nonlineq:static})
for $\Delta_{0}=0, N_{f}=2$: the bold (black) line for $\lambda=0.3$,  the dashed
(red) line for $\lambda=0.27$, and the dash-dotted (green) line for $\lambda=0.25$.
}
\label{gap-dependence}
\end{figure}

The above analysis is adequate precisely at the critical coupling, i.e.,
at the bifurcation point of the original nonlinear equation. To study
momentum dependence of solutions of Eq.(\ref{gap:eq}) beyond the critical
point  we now turn back to the case of static polarization when $J=\pi/(1+g)$
and Eq.(\ref{gap:eq}) is written in the form
\be
\Delta(p)=\Delta_{0}+\frac{2\lambda}{\pi}\int\limits_{0}^{\Lambda}\frac{dk\,k \Delta(k)}
{\sqrt{k^{2}+\Delta^{2}(k)}}{\cal K}(p,k),\quad \lambda=\frac{\alpha}{2(1+\pi N_{f}\alpha/4)},
\label{nonlineq:static}
\ee
with the kernel (compare with Eq.(\ref{kernel-identity}))
\be
{\cal K}(p,k)=\frac{1}{p+k}K\left(\frac{2\sqrt{p\,k}}{p+k}\right)=
\frac{\theta(p-k)}{p}K\left(\frac{k}{p}\right)
+\frac{\theta(k-p)}{k}K\left(\frac{p}{k}\right).
\label{nonlineq:kernel}
\ee
The gap equation is essentially different from a gap equation in BCS theory
where the gap is momentum independent. In Fig.\ref{gap-dependence}, we present the results
of our numerical solution to Eq.(\ref{nonlineq:static}) for $\Delta_{0}=0$, $N_{f}=2$
and several values of $\lambda$.
The gap is weakly dependent on a momentum up to values $p\sim \Delta(0)$ after which the behavior
becomes more steep. To estimate the  gap $\Delta(0)$ we need to know the bandwidth
parameter $\Lambda$ which can be obtained by equating the wave vector
integral over the Brillouin zone to the integral over two Dirac points with a cutoff at $k_{c}$.
We get $k_{c}=(\pi/\sqrt{3})^{1/2}(2/a)$ where $a$ is the lattice
constant, therefore, restoring $\hbar$ and $v_{F}=\sqrt{3}t a/(2\hbar)$, we find
$\Lambda=\hbar v_{F}k_{c}=\sqrt{\pi\sqrt{3}}t\approx2.33t$.
For  the hopping parameter $t=3\,\mbox{eV}$, we obtain $\Lambda\approx7\mbox{eV}$.
The maximal possible gap $\Delta(0)$
is reached for $\alpha\to\infty$ that corresponds to the value $\lambda=1/\pi\approx0.32$ ($N_{f}=2$).
For the values of $\lambda$'s used in Fig. \ref{gap-dependence} we find the estimates:
$\Delta(0)=200\, \mbox{K}, 40\, \mbox{K}, 25 \,\mbox{K}$ for $\lambda=0.3, 0.27, 0.25$, respectively.

To get insight into analytical solutions of Eq.(\ref{nonlineq:static}) we
approximate the elliptic integral functions in (\ref{nonlineq:kernel}) by their
asymptotical values at $p\ll k$ and $p\gg k$, we obtain the kernel
\be
{\cal K}(p,k)=\frac{\pi}{2}\left(\frac{\theta(p-k)}{p}+\frac{\theta(k-p)}{k}\right).
\label{kernel-approximation}
\ee
This allows us to reduce the nonlinear integral equation (\ref{nonlineq:static})  to the second
order nonlinear differential equation
\be
\left(p^{2}\Delta^{\prime}(p)\right)^{\prime}+\lambda\frac{p\Delta(p)}{\sqrt{p^{2}+
\Delta^{2}(p)}}=0,
\label{nonlinear-diff-eq}
\ee
with the infrared (IR) and ultraviolet (UV) boundary conditions:
\ba
&& p^{2}\Delta^{\prime}(p)\Big|_{p=0}=0,
\label{IRBC}\\
&& \left(p\Delta(p)\right)^{\prime}\Big|_{p=\Lambda}=\Delta_{0}.
\label{UVBC}
\ea
Eq.(\ref{nonlinear-diff-eq}) is scale invariant, i.e., if $\Delta(p)$ is a solution,
then $l\Delta(p/l)$ is also a solution. The scale invariance is broken by the UV boundary
condition only.

The chiral condensate $\langle0|\overline\Psi\Psi|0\rangle$  is the order parameter
for the semimetal-insulator transition in graphene, it breaks spontaneously
the initial $U(2N_{f})$ symmetry down the $U(N_{f})\otimes U(N_{f})$ but keeps
parity and time-reversal invariances. It is expressed through the full fermion
propagator as follows
\ba
\langle0|\overline\Psi\Psi|0\rangle&=&-\lim\limits_{x\to0}\langle0|T\Psi(x)\overline\Psi(0)|0\rangle
=-i{\rm tr}\int\limits_{-\infty}^{\infty}\frac{d\omega}{2\pi}\int^{\Lambda}\frac{d^{2}p}{(2\pi)^{2}}
G(\omega,\mathbf{p})\nonumber\\
&=&-\frac{N_{f}}{\pi}\int\limits_{0}^{\Lambda}\frac{dpp\Delta(p)}{\sqrt{p^{2}
+\Delta^{2}(p)}}=\frac{N_{f}}{\pi\lambda}p^{2}\Delta^{\prime}(p)\Big|_{p=\Lambda},
\label{condensate-through-gap}
\ea
where for the last equality we used Eq.(\ref{nonlinear-diff-eq}). Hence the condensate is
nontrivial if a nontrivial solution of the gap equation exists.

One can easily find the solutions of Eq.(\ref{nonlinear-diff-eq}) in two asymptotic regions.
For $p\ll\Delta(p)$,
\be
\Delta(p)=C_{1}+\frac{C_{2}}{p}.
\ee
The IR boundary condition (\ref{IRBC}) implies  $C_{2}=0$, therefore, $\Delta(p)\simeq C_{1}$
for $p\ll\Delta(p)$. For $p\gg\Delta(p)$,
\be
\Delta(p)\simeq C_{3}p^{-\gamma_{+}}+C_{4}p^{-\gamma_{-}}, \quad \gamma_{\pm}=\frac{1}{2}
\pm\sqrt{\lambda-\lambda_{c}}.
\label{asymptotic}
\ee
Clearly, in order to find a solution of Eq.(\ref{gap:eq}) one needs to show that there
exists a solution of the nonlinear differential equation
(\ref{nonlinear-diff-eq}) which connects the asymptotic $\Delta(p)\simeq const$
in the infrared region, $p\to0$, with the asymptotic (\ref{asymptotic}) at
large momenta. For this, let us define
\be
\Delta(p)=e^{t}u(t+t_{0}), \quad t=\ln p,
\label{definition-u(t)}
\ee
then the function $u(t)$ satisfies the differential equation
\be
u^{\prime\prime}+3u^{\prime}+2u+\lambda\frac{u}{\sqrt{1+u^{2}}}=0.
\label{u-eq}
\ee
The IR boundary condition implies
\be
e^{2t}\left(u^{\prime}+u\right)\Big|_{t=-\infty}=0.
\label{IRBC-u}
\ee
We require that $e^{t}u(t)\to1$ as $t\to-\infty$ since all other solutions for $\Delta(p)$
are obtained by varying the constant $t_{0}$. For this
normalization, the infrared scale for the general solution is given by
$\Delta(0)=e^{-t_{0}}$.

The dependence of the integral equation (\ref{gap:eq}) on the bare
gap $\Delta_{0}$ now becomes an
ultraviolet boundary condition for the differential equation, it is
\be
u^{\prime}\left(t_{\Lambda}+t_{0}\right)+2u\left(t_{\Lambda}+t_{0}\right)=\Delta_{0}/\Lambda.
\label{UVBC-u}
\ee
This condition determines the value of the parameter $t_{0}=-\ln\Delta(0)$ as a function
of the coupling constant, $\lambda$, the bare gap, $\Delta_{0}$, and the cutoff, $\Lambda$.
Eq.(\ref{u-eq}) can be rewritten in the form
\be
u^{\prime\prime}+3u^{\prime}=-\frac{d}{d u}V(u),\quad V(u)=u^{2}+\lambda\sqrt{1+u^{2}}.
\label{particle-equation}
\ee
or, equivalently,
\be
\left(\frac{1}{2}(u^{\prime})^{2}+V(u)\right)^{\prime}=-3(u^{\prime})^{2}.
\ee
Eq.(\ref{particle-equation}) is the equation for a particle of unit mass moving in
a potential $V$ with friction proportional to velocity. The ``energy''
$\frac{1}{2}(u^{\prime})^{2}+V(u)$  reaches its absolute minimum at $u=0$, hence
the particle moves toward $u=0$ damped by the friction. The asymptotical behavior near $u=0$
is described by the linearized equation
\be
u^{\prime\prime}+3u^{\prime}+(2+\lambda)u=0,
\ee
and depends on whether the coupling $\lambda>\lambda_{c}\equiv 1/4$, or $\lambda<\lambda_{c}$,
\ba
&&u(t)\rightarrow \frac{B}{\sqrt{\lambda_{c}-\lambda}}\,e^{-3t/2}\sinh\left[\sqrt{\lambda_{c}-\lambda}
\,(t +\delta)\right],\quad t\rightarrow\infty,\quad
\mbox{weak coupling}\,\, (\lambda<\lambda_{c}),
\label{UV-weakcoupling}\\
&&u(t)\rightarrow \frac{A}{\sqrt{\lambda-\lambda_{c}}}\,e^{-3t/2}\sin\left[\sqrt{\lambda-\lambda_{c}}
\,(t +\delta)\right],\quad t\rightarrow\infty,\quad \mbox{strong coupling}\,\,\lambda\ge\lambda_{c},
\label{UV-strongcoupling}
\ea
where the constants $A,B$, and $\delta$ are functions of the coupling constant $\lambda$.
We explicitly singled out the factor
$1/\sqrt{\lambda_{c}-\lambda}$ in front of Eqs.(\ref{UV-weakcoupling}),
(\ref{UV-strongcoupling}) since the function $u(t)$ must be nontrivial at
$\lambda=\lambda_{c}$. Obviously,
$A(\lambda=\lambda_{c})=B(\lambda=\lambda_{c})$.

The asymptotics (\ref{UV-weakcoupling}) and (\ref{UV-strongcoupling})
imply that at weak coupling the particle situated initially at $u(-\infty)$
reaches $u=0$ for infinite time, meanwhile, at strong coupling it will get to
$u=0$ in a finite time and then oscillate there with damped amplitude.

It is easy to see that at weak coupling there are no nontrivial solutions satisfying
the UV boundary condition for $\Delta_{0}=0$.
For strong coupling, the UV boundary condition (\ref{UVBC-u}) with $\Delta_{0}=0$
admits an infinite number of solutions for the gap scale $\Delta(0)$, corresponding to
different solutions of the equation
\ba
u^{\prime}\left(t_{\Lambda}+t_{0}\right)+2u\left(t_{\Lambda}+t_{0}\right)\approx
\frac{A\sqrt{\lambda}}{\sqrt{\lambda-\lambda_{c}}}\,e^{-3(t_{\Lambda}+t_{0})/2}
\sin\left(\theta+\phi\right)=0,
\ea
where
\be
\theta=\sqrt{\lambda-\lambda_{c}}\left(t_{\Lambda}+t_{0}+\delta\right)=\sqrt{\lambda-\lambda_{c}}
\ln\left(\frac{e^{\delta}\Lambda}{\Delta(0)}\right),\quad \phi=\arctan\left(2\sqrt{\lambda-\lambda_{c}}\right).
\ee
Hence, the solution is given by $\theta=\pi n-\phi$, or
\be
\Delta(0)=\Lambda e^{\delta}\exp\left(-\frac{\pi n-\phi}{\sqrt{\lambda-\lambda_{c}}}\right),
\quad n=1,2,\dots.
\label{ne-solution}
\ee
The solution without nodes, $n=1$, corresponds to the ground state  since it
generates the largest fermion gap and has the lowest energy.
The critical coupling $\lambda_{c}=1/4$ is a bifurcation point of the integral equation
(\ref{gap:eq}) with the static vacuum polarization \cite{footnote2}.
The expression (\ref{ne-solution}) for the gap implies that this bifurcation
point corresponds to a continuous phase transition of infinite order.
 As was shown in Ref.\cite{excitonic-instability}, the critical coupling
$\lambda_{c}$ is closely related to the phenomenon ``fall into the center'' in quantum mechanics problem.
A similar situation takes place in the strong coupling QED4 \cite{review} where
in the ladder approximation (and more generally in quenched approximation when fermions loops
are neglected \cite{Holdom}) the phase transition is also of infinite order. The dimensionless
correlation length,
\be
\xi=\frac{\Lambda}{\Delta(0)}\sim\exp\left(\frac{\pi}{\sqrt{\lambda-\lambda_{c}}}\right),
\label{BKT-corrlength}
\ee
exponentially grows when $\lambda\to\lambda_{c}$. Such a behavior is inherent for the
Berezinskii--Kosterlitz--Thouless phase transition (or the conformal phase
transition (CPT) \cite{CPT}) and, obviously, is related to the scale invariance
of the problem under consideration. Note, however, that taking into account the
finite size of graphene samples should turn this phase transition into a second
order one (as it is shown for QED3 in Ref.\cite{GusyninReenders}).

Eq.(\ref{BKT-corrlength}) means also that in nonperturbative phase
there is a nontrivial running of the coupling $\alpha$ (or the Fermi velocity $v_{F}$)
though we neglected its perturbative running. Defining the $\beta$ function in
a standard way, we find
\be
\beta({\alpha})\equiv\Lambda\frac{d\alpha}{d\Lambda}= -\frac{\pi}{4}(1+\pi N_{f}\alpha/4)^{2}
(\lambda-\lambda_{c})^{3/2},\quad\lambda>\lambda_{c},
\label{beta(alpha)}
\ee
where $\lambda$ is defined in Eq.(\ref{nonlineq:static}).
The $\beta$ function depends nonanalytically on the coupling $\alpha$ and can not
be obtained in perturbation theory. We expect that if a perturbative running of $\alpha$
is included the critical point $\lambda_{c}$ becomes a second order phase transition
point  on which the $\beta$ function (\ref{beta(alpha)}) is continuous
when approached from both perturbative and nonperturbative phases.

The order parameter $\langle\overline\Psi\Psi\rangle$ in terms of the function $u(t)$ is given by
\ba
\langle\overline\Psi\Psi\rangle=\frac{N_{f}}{\pi\lambda}e^{2t_{\Lambda}}\left(u^{\prime}(t_{\Lambda}+t_{0})+
u(t_{\Lambda}+t_{0})\right),
\label{barpsipsi-through-u}
\ea
and equals
\ba
\langle\overline\Psi\Psi\rangle=-\frac{N_{f}A}{\pi\sqrt{\lambda(\lambda-\lambda_{c})}}\,
\Lambda^{1/2}\Delta^{3/2}(0)\sin(2\phi)
=-\frac{N_{f}A}{\pi{\lambda}^{3/2}}\Lambda^{1/2}\Delta^{3/2}(0),
\ea
where the relation $\theta=\pi-\phi$ was used.

For nonzero bare gap, $\Delta_{0}\neq0$, we obtain the following equation for the scale $\Delta(0)$:
\be
\Delta_{0}=\frac{A\sqrt{\lambda}}{\sqrt{\lambda-\lambda_{c}}}\,\frac{\Delta^{3/2}(0)}{\sqrt{\Lambda}}
\sin\left(\theta+\phi\right),
\label{Delta0-versus-Delta}
\ee
and  the order parameter,
\be
\langle\overline\Psi\Psi\rangle=\frac{N_{f}}{\pi\lambda}\Lambda\left[\Delta_{0}-
\frac{A}{\sqrt{\lambda-\lambda_{c}}}\frac{\Delta^{3/2}(0)}{\sqrt{\Lambda}}\sin\theta\right].
\ee
Let us write $\theta+\phi=\pi-\epsilon$ where $\epsilon$ tends to zero when $\Delta_{0}\to0$
and $\lambda\to\lambda_{c}$.
Then the above equations are rewritten as
\ba
&&\Delta_{0}=\frac{A\sqrt{\lambda}}{\sqrt{\lambda-\lambda_{c}}}\,
\frac{\Delta^{3/2}(0)}{\sqrt{\Lambda}}\sin\epsilon,\\
&&\langle\overline\Psi\Psi\rangle=\frac{N_{f}\Lambda}{\pi\lambda}\left[\frac{2\lambda-1}{2\lambda}
\Delta_{0}-A\frac{\Delta^{3/2}(0)}{\sqrt{\Lambda}}\frac{\cos\epsilon}{\sqrt{\lambda}}
\right].
\label{order-parameter}
\ea
In such a form the equations are convenient for finding critical exponents near the
phase transition point $\lambda_{c}$. Critical exponents describe the approach to criticality
of such quantities as the correlation length, the order parameter, the susceptibility, etc.,
they are defined  in a standard way \cite{Miranskybook,Ma}:
\ba
&&\xi=\frac{\Lambda}{\Delta(0)}\sim\left(\lambda-\lambda_{c}\right)^{-\nu},\quad
\frac{\langle\overline\Psi\Psi\rangle}{\Lambda^{2}}\sim\left(\lambda-\lambda_{c}\right)^{\beta},\quad
\chi=\frac{\partial\langle\overline\Psi\Psi\rangle}{\partial\Delta_{0}}\Big|_{\Delta_{0}=0}
\sim\left(\lambda-\lambda_{c}\right)^{-\gamma},\quad \lambda\to\lambda_{c},\\
&&\langle\overline\Psi\Psi\rangle\Big|_{\lambda=\lambda_{c}}\sim \Delta_{0}^{1/\delta},\quad \Delta_{0}
\to0.
\ea
If the theory of second order phase transition is applicable, then the exponents are assumed to
obey the following hyperscaling relations in spaces of arbitrary dimension
$D$:
\ba
2\beta+\gamma=D\nu,\quad 2\beta\delta-\gamma=D\nu,\quad \frac{\delta-1}{\delta+1}=\frac{2-\eta}{D},
\quad \beta=\nu\frac{D-2+\eta}{2}.
\label{hyperscaling}
\ea
Here the exponent $\eta$ describes the behavior of the correlation function
\be
\langle\overline\Psi\Psi(r)\overline\Psi\Psi(0)\rangle\Big|_{\lambda=\lambda_{c}}
\propto r^{-D+2-\eta},\quad r\to\infty.
\label{correlator}
\ee
Using Eq.(\ref{order-parameter}), we find
\be
\langle\overline\Psi\Psi\rangle\Big|_{\lambda=\lambda_{c}}=-\frac{4N_{f}\Lambda}{\pi}
\left[\Delta_{0}+\frac{2A\Delta^{3/2}(0)}{\sqrt{\Lambda}}\right],
\label{orderparameter}
\ee
and due to Eq.(\ref{Delta0-versus-Delta}),
\be
\Delta(0)\sim\left(\frac{\Delta_{0}}{\ln(\Lambda/\Delta_{0})}\right)^{2/3},\quad \lambda=\lambda_{c},
\ee
the critical exponent $\delta=1$, and from hyperscaling relations we obtain
\be
\eta=2,\quad \gamma=0,\quad \beta=\frac{3\nu}{2}.
\ee
The infinite order phase transition with the correlation length (\ref{BKT-corrlength})
formally corresponds to the limit
\be
\beta=\frac{3\nu}{2}\to\infty.
\ee
Certainly, the infinite order phase transition is quite different from that one
studied in lattice simulations \cite{Drut} where a second order phase transition was found
with the critical exponents $\delta\sim2.3, \beta\sim0.8,
\gamma\sim1 (N_{f}=2)$. One of the reasons for such a difference might be a finite size
of a lattice that changes the kind of phase transition.
Effectively, the finite size of a lattice can be taken into account by introducing
an infrared cutoff ($k_{0}\sim \pi/L$, $L$ is a linear size of the system) in the integral
equation (\ref{gap:eq}) \cite{GusyninReenders,Fisher}.
Another reason could be that one should take into account residual lattice interactions,
i.e., the present analysis has to be further refined by incorporating
 effective four-fermion terms (see the next section).

\section{PHASE DIAGRAM IN THE MODEL WITH ADDITIONAL FOUR-FERMION INTERACTION}
\label{phase-diagram}

As discussed in Introduction, when comparing the results of lattice simulations \cite{Drut}
with analytical calculations one should have in mind that  the continuum theory
described by Lagrangian (\ref{action}) putted on a lattice  contains
unavoidably additional interaction terms, in particular, local four-fermion interaction
terms. This means that it would be appropriate  to add to the continuum theory some
local four-fermion interaction terms in addition to the long range
Coulomb interaction in Eq.(\ref{action}).
 The amount of induced couplings depends of course on the particular
lattice regularization employed. Furthermore, according to \cite{Alicea,Herbut,Aleiner},
the effective continuum model for quasiparticles in graphene in addition to the Coulomb
interaction should contain contact four-fermion interaction
terms which arise from the original lattice tight-binding model. Usually these
residual four-fermion terms are irrelevant operators from the point of view of the
renormalization group, however, we will show that they can play a significant role in
the critical behavior. In order to study how these
four-fermion terms influence the gap generation, we will consider in this section
a continuum model with the Coulomb interaction and the Gross-Neveu four-fermion
interaction of the type
\be
{\cal L}_{4}=\frac{G}{2}(\overline\Psi_{a}\Psi_{a})^{2},
\label{4fermion-int}
\ee
where the four-fermion coupling constant $G$ is of the order of the lattice
constant and the ``flavor'' index $a=1,2,\dots,N_{f}$, $N_{f}=2$ for physical spin-$1/2$ electrons.
 The interaction term (\ref{4fermion-int}) breaks the
initial $U(2N_{f})$ symmetry of the action
(\ref{action})  down to the $U(N_{f})\otimes U(N_{f})\otimes Z_{2}$ symmetry.
While the gap term $\Delta\overline\Psi\Psi$ is invariant under the $U(N_{f})\otimes U(N_{f})$,
it is not under the discrete chiral $Z_{2}$ - symmetry: $\Psi\rightarrow\gamma_{5}
\Psi, \overline\Psi\rightarrow-\overline\Psi\gamma_{5}$. In the absence of the bare gap term, $\Delta_{0}
\overline\Psi\Psi$, the $Z_{2}$ symmetry  forbids the fermion gap
generation in perturbation theory. The appearance of the energy gap is due to the
spontaneous breaking of the above discrete chiral symmetry that leads to a neutral condensate
$\langle\overline\Psi\Psi\rangle$ of fermion-antifermion pairs (excitonic condensate).

The gap equation (\ref{gap:eq}) is modified in the presence of the interaction
(\ref{4fermion-int}) in the following way:
\be
\Delta(p)=\Delta_{0}-G(1-\frac{1}{4N_{f}})\langle\overline\Psi\Psi\rangle+\frac{\alpha}{\pi^{2}}
\int\limits_{0}^{\Lambda}
\frac{d k\,k\,\Delta(k)}{\sqrt{k^{2}+\Delta^{2}(k)}}
\,{\cal K}(p,k),
\ee
where the condensate $\langle\overline\Psi\Psi\rangle$ contributes like a bare fermion gap
and can be computed from the fermion self-energy; the factor $1-1/4N_{f}$ in the second
term on the right hand side takes into account both Hartree and Fock  ($- 1/4N_{f}$) contributions.
For the sake of simplicity we consider only the Hartree term, if necessary, the Fock contribution
can be easily restored in final formulas \cite{Fock-term}.
In the approximation to the kernel used above (Eq.(\ref{kernel-approximation})) the condensate is
given by the expression (\ref{condensate-through-gap}). The condensate does not change
the differential equation (\ref{nonlinear-diff-eq}), however,  it modifies the ultraviolet
boundary condition (\ref{UVBC}):
\be
\left[\left(1+\frac{\tilde{g}N_{f}}{\lambda}\right)p\Delta^{\prime}(p)+\Delta(p)\right]
\Big|_{p=\Lambda}=\Delta_{0},
\label{EoS}
\ee
where we introduced the notation $\tilde{g}=G\Lambda/\pi$ and $\lambda$
is defined in Eq.(\ref{nonlinear-diff-eq}). Using the definition of the gap function
in terms of the $u(t)$ function (\ref{definition-u(t)}) and the asymptotic behavior of the last one,
Eqs.(\ref{UV-weakcoupling}),(\ref{UV-strongcoupling}), the equation (\ref{EoS}) can be written
for $\Lambda\gg \Delta(0)$ in the following form:
\ba
&&B\frac{\Delta^{3/2}(0)}{\sqrt{\Lambda}}\left[\left(1+\frac{\tilde{g}N_{f}}{\lambda}\right)\cosh\left(
\omega\ln\frac{\Lambda e^{\delta}}{\Delta(0)}\right)
+\frac{1-\tilde{g}N_{f}/\lambda}{2\omega}\sinh\left(
\omega\ln\frac{\Lambda e^{\delta}}{\Delta(0)}\right)\right]=\Delta_{0},\,\,
\omega=\sqrt{\lambda_{c}-\lambda},\,\lambda<\lambda_{c},
\label{EoS-subcritic-lambda}\\
&&B\frac{\Delta^{3/2}(0)}{\sqrt{\Lambda}}\left[1+\frac{\tilde{g}N_{f}}{\lambda}
+\frac{1-\tilde{g}N_{f}/\lambda}{2}\ln\frac{\Lambda e^{\delta}}{\Delta(0)}\right]=\Delta_{0},\quad
\lambda=\lambda_{c},\\
&&A\frac{\Delta^{3/2}(0)}{\sqrt{\Lambda}}\left[\left(1+\frac{\tilde{g}N_{f}}{\lambda}\right)\cos\left(
\tilde{\omega}\ln\frac{\Lambda e^{\delta}}{\Delta(0)}\right)
+\frac{1-\tilde{g}N_{f}/\lambda}{2\tilde{\omega}}\sin\left(
\tilde{\omega}\ln\frac{\Lambda e^{\delta}}{\Delta(0)}\right)\right]=\Delta_{0},\quad
\tilde{\omega}=\sqrt{\lambda-\lambda_{c}},\, \lambda>\lambda_{c},
\ea
and we remind that in the utilized approximation $\lambda_{c}=1/4$.
These equations imply the following solutions for the dynamical gap
in the case $\Delta_{0}=0$:
\ba
&&\Delta(0)=\Lambda e^{\delta}\left[\frac{\tilde{g}N_{f}(1-2\omega)-\lambda(1+2\omega)}{\tilde{g}N_{f}(1+2\omega)
-\lambda(1-2\omega)}\right]^{\frac{1}{2\omega}},\quad \lambda<\lambda_{c},
\label{Delta-subcritical}\\
&&\Delta(0)=\Lambda e^{\delta}\exp\left[-2\frac{\tilde{g}N_{f}+1/4}{\tilde{g}N_{f}-1/4}\right],\quad
\lambda=\lambda_{c}=\frac{1}{4},\quad \tilde{g}N_{f}>\frac{1}{4}
\label{Delta-crit1}\\
&&\Delta(0)=\Lambda e^{\delta}\exp\left[-\frac{\pi n}{\tilde{\omega}}-\frac{1}{\tilde{\omega}}
\arctan\left(2\tilde{\omega}\frac{\tilde{g}N_{f}+\lambda}{\tilde{g}N_{f}-\lambda}\right)\right],
\quad\lambda>\lambda_{c},
\quad n=1,2,\dots.
\label{Delta-crit2}
\ea
Note that the solutions (\ref{Delta-crit1}), (\ref{Delta-crit2}) contain essential
singularity, the first solution at $\tilde{g}=1/4N_{f}$ and the second one at $\lambda=\lambda_{c}$.

\begin{figure}[ht]
\includegraphics[width=8.5cm]{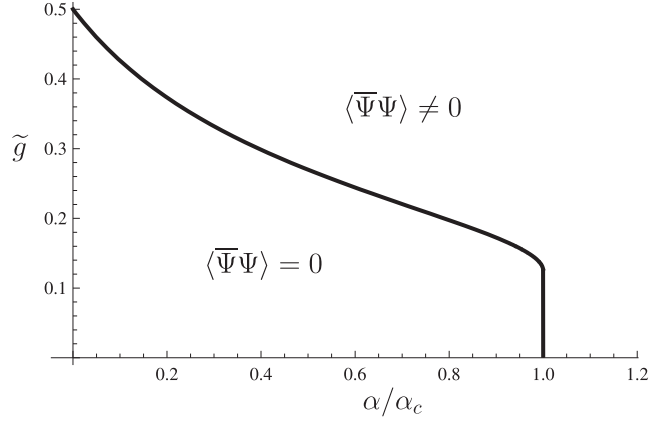}
\caption{Phase diagram for $N_{f}=2$.
}
\label{phase}
\end{figure}

Setting $\Delta(0)=0$, we find the critical line separating the spontaneously broken and
unbroken phases of the chiral symmetry:
\ba
\left\{\begin{array}{c}\tilde{g}=\frac{1}{4N_{f}}\left(1+\sqrt{1-\lambda/\lambda_{c}}\right)^{2},
\quad \mbox{for}\quad \lambda\le\lambda_{c}=\frac{1}{4},\\
\tilde{g}<\frac{1}{4N_{f}},\quad \lambda=\lambda_{c}=\frac{1}{4}.\end{array}\right.
\label{phasediagram}
\ea
The phase diagram in the plane of two coupling constants  is displayed in Fig.\ref{phase} \cite{Fock-term}.
Above the critical line the gap equation for the fermion self-energy $\Delta(p)$ has a nontrivial
solution. Thus, the chiral symmetry is dynamically broken that implies the existence of
a nonzero vacuum condensate $\langle\overline\Psi\Psi\rangle$.
For $\tilde{g}=0$, the condition for a gap generation becomes $\lambda>\lambda_{c}$ and the corresponding
critical coupling coincides with Eq.(\ref{critline-Coulomb}).
On the other hand, in the other limiting case
$\alpha=0$, $\tilde{g}_{c}=1/N_{f}$ coincides with the critical coupling in the
Gross--Neveu model. In the part of the phase diagram above the critical line $\lambda<\lambda_{c}$
the short-range four-fermion interactions play important role for the condensate formation,
meanwhile, in the region $\lambda>\lambda_{c}$ Coulomb forces are mainly responsible for
the condensate formation.

We consider now the phase transition along the upper part of the critical line
and compute the critical exponents.
Since we consider non-running coupling $\alpha$ (in absence of running of the Fermi
velocity $v_{F}$) the renormalization group flow can be determined from Eq.(\ref{Delta-subcritical})
which near a critical point takes the form
\be
\frac{\Delta(0)}{\Lambda}\sim\left(\frac{\tilde{g}-\tilde{g}_{1}}{\tilde{g}-\tilde{g}_{2}}\
\right)^{\frac{1}{2\omega}},\quad\tilde{g}_{1}=\frac{(1+2\omega)^{2}}{4N_{f}},\quad
\tilde{g}_{2}=\frac{(1-2\omega)^{2}}{4N_{f}},\quad\tilde{g}>\tilde{g}_{1}>\tilde{g}_{2}.
\label{nearcriticalDelta}
\ee
It implies an explicit form of the $\beta$ function for the coupling $\tilde{g}$:
\be
\beta(\tilde{g},\alpha)\equiv\Lambda\frac{\partial \tilde{g}}{\partial\Lambda}\Big|_{\alpha,\Delta(0)}
=-N_{f}(\tilde{g}-\tilde{g}_{1})(\tilde{g}-\tilde{g}_{2}),\quad \tilde{g}>\tilde{g}_{1}.
\label{beta-function}
\ee
Eq.(\ref{beta-function}) has indeed a nontrivial fixed line at $\tilde{g}=\tilde{g}_{1}$.
We stress that the $\beta$ function (\ref{beta-function}) is obtained in nonperturbative
phase where a quasiparticle gap is spontaneously generated. In perturbative
phase the $\beta$ function was calculated in \cite{Juricic} (see, Eq.(7) there), in the leading
order in $1/N_{f}$ and small coupling $\alpha$ both $\beta$ functions behave as $\beta\simeq
-(g-g_{0})$ near the fixed point $g_{0}\sim (1-\alpha)/N_{f}$.
As is seen from Eq.(\ref{nearcriticalDelta}), the phase transition  is of the second order.
Denoting the deviation from the critical line as $\tau\equiv \tilde{g}-\tilde{g}_{1}$
and because $\Delta(0)\sim\tau^{1/2\omega}$ ($\tau\to0$) we find the exponent
\be
\nu=\frac{1}{2\omega}.
\label{nu-exponent}
\ee
The condensate is given by
\be
\langle\overline\Psi\Psi\rangle=\frac{N_{f}B}{\pi\lambda}\Delta^{3/2}(0)\Lambda^{1/2}
\left[\cosh\left(\omega\ln\frac{\Lambda e^{\delta}}{\Delta(0)}\right)-\frac{1}{2\omega}
\sinh\left(\omega\ln\frac{\Lambda e^{\delta}}{\Delta(0)}\right)\right].
\label{condensate-subcritical}
\ee
On the critical line,  Eq.(\ref{EoS-subcritic-lambda}) implies
 $\Delta(0)\sim\Delta^{1/(3/2 +\omega)}_{0}$. Substituting it into
the expression for the fermion condensate gives the critical scaling relation
\be
\langle\overline\Psi\Psi\rangle\sim\Delta^{\frac{3}{2}-\omega}
\sim\Delta_{0}^{(3/2 -\omega)/(3/2 +\omega)},
\ee
thus the critical exponent
\be
\delta=\frac{3/2 +\omega}{3/2 -\omega}.
\label{delta-exponent}
\ee
It is equal to $\delta=2$ in the case of the Gross--Neveu model when $\alpha=0$
and $\delta\to1$ for $\alpha \to \alpha_c$. It is easy to find also the $\beta$ exponent:
\be
\beta=\frac{1}{2\omega}\left(\frac{3}{2}-\omega\right).
\label{beta-exponent}
\ee
Finally, it follows from Eqs.(\ref{EoS-subcritic-lambda}), (\ref{condensate-subcritical}) that
\be
\partial_{\Delta_{0}}\langle\overline\Psi\Psi\rangle\Big|_{\Delta_{0}=0}\sim\tau^{-1},
\quad\tau\to0,
\ee
hence the exponent $\gamma=1$. The found critical exponents satisfy the hyperscaling
relations (\ref{hyperscaling}). The additional critical exponent $\eta$ may be calculated
independently, or using hyperscaling relations, $\eta=2-2\omega$.
By definition, the anomalous dimension $\gamma_{m}$ of the composite operator is given by
\mbox{dim}($\overline\Psi\Psi)=D-1-\gamma_{m}$, then  the correlator (\ref{correlator})
implies the relation $\eta=D-2\gamma_{m}$. In our case,
$D=3$, we obtain $\gamma_{m}=1/2 +\omega$. The dynamical dimension of the four-fermion
interaction term equals \mbox{dim}$(\overline\Psi\Psi)^{2}=2$\mbox{dim}
$(\overline\Psi\Psi)=4-2\gamma_{m}$. Because $1/2\le \gamma_{m}\le 1$ along the
critical line, \mbox{dim}$(\overline\Psi\Psi)^{2}\le3$ and
the four-fermion operator $(\overline\Psi\Psi)^{2}$ acts as a renormalizable one. In the
renormalization group terminology,
the $(\overline\Psi\Psi)^{2}$ becomes a relevant operator in the scaling region while it is irrelevant
away from the critical line, in accordance with standard renormalization group approach \cite{Herbut},
 as its effects are suppressed by powers of cutoff.
On the other hand, the anomalous dimension $\gamma_{m}$  governs the behavior of
the amputated Bethe-Salpeter wave function (form factor) of bound
states, $\chi^{(amp)}(q)\sim (q/\Delta(0))^{\gamma_{m}-1}$, in the range of momenta
$\Delta(0)\ll q\ll\Lambda$. The ``critical'' value $\gamma_{m}=1/2$
separates loose ($\gamma_{m}<1/2$) and tight ($\gamma_{m}\ge1/2$) bound states.
The wave functions with large $\gamma_{m}\, (\gamma_{m}>1/2)$ slowly decrease with
 momentum, they describe tight bound states which  are relevant for critical
sca1ing laws of a theory \cite{Miransky}. Since such bound states resemble
point-like particles,  the scaling properties of a theory
can be described by an effective Lagrangian with elementary scalar fields
(for recent such an approach, see, Ref.\cite{HerbutVafek}).
The computer simulations of lattice graphene model may reveal in principle
the existence of such tight bound states.

We see that the additional Gross--Neveu four-fermion interaction
plays an important role: First, it changes the order of a phase transition from
the infinite to the second order one. Second, the critical coupling
becomes lower than in the model with the pure Coulomb interaction. Third, the
critical exponents stay closer to those obtained in lattice simulations
\cite{Drut}. Further, the critical indices depend on the coupling $\alpha$
along the critical line $0<\alpha<\alpha_{c}$ and satisfy the hyperscaling
relations. The phase diagram (\ref{phasediagram}) resembles closely those
obtained in the strong coupling QED4 \cite{critlineQED4} and QED3
\cite{critlineQED3}.
Since the phase transition is of second order along the
$0<\alpha<\alpha_{c}$ part  of the critical curve, Eq.(\ref{phasediagram}),
resonances should exist on the symmetric side of the curve, whose masses tend to
zero as the critical curve is approached \cite{Appelquist,GRH}. The part of the critical curve
with $\tilde{g}<1/4N_{f}$ is rather special and is related to the conformal phase transition \cite{CPT}.
It is characterized by a gap function having an essential singularity at the transition point, and by
abrupt change of the spectrum of light excitations as the critical point is crossed:
light bound states near the critical line are absent in the
symmetric phase, however, they are present in the phase with broken symmetry
(for a discussion in detail of the CPT in QED3, see
Ref.\cite{Appelquist,Shpagin}). The corresponding effective potential for the
order parameter $\langle \overline\Psi\Psi \rangle$, unlike the
familiar Ginzburg--Landau potential, was shown to have a branched fractal
structure in the region $\alpha>\alpha_{c}$, where the Coulomb interaction is
mainly responsible for the bound states formation \cite{GGMS}.

We hope that at least some features of the picture outlined above
will be confirmed in experiments with suspended clean graphene.

\section{Conclusions}

In this paper we studied the gap generation in suspended clean graphene at neutral point. Solving
the Schwinger-Dyson equation with the frequency dependent polarization function
we found analytically that the critical coupling constant for onset of a gap equals
$\alpha_{c}=0.92$ which is close to the value obtained in Monte Carlo simulations. We showed
that the critical coupling $\alpha_c$ corresponds to the infinite order phase transition
in the case of purely Coulomb interaction with peculiar critical
exponents while Monte Carlo simulations point to the second order phase transition with
different critical exponents.

Adding the Gross-Neveu four-fermion interaction that is present in
the continuum limit of the lattice model we found the critical line
Eq.(\ref{phasediagram})  in the plane of Coulomb
and four-fermion coupling constants separating zero-gap and gapped phases.
We showed that the order of a phase transition changes from the
infinite to the second order one along the part $0<\alpha<\alpha_{c}$ of the
critical line and the critical coupling becomes lower than in the model with
pure Coulomb interaction. The critical exponents $\nu,\delta,\beta$ along the line of
second order phase transition are given by the expressions (\ref{nu-exponent}),
(\ref{delta-exponent}), (\ref{beta-exponent}), respectively, and the exponent $\gamma=1$.
These exponents satisfy hyperscaling relations and characterize the transition between
phases with distinct symmetry properties and become, in general, functions of
the Coulomb coupling $\alpha$, or the four-fermion coupling $\tilde{g}$.
They are close to the critical exponents obtained in lattice simulations \cite{Drut}.

The other part of the critical curve
with $\tilde{g}<1/4N_{f}$ is rather special and is related to the conformal phase transition
characterized by an essential singularity at the transition point and by
abrupt change of the spectrum of light excitations as the critical point is crossed.
However, the shape of the last part of phase transition curve might be strongly influenced
by the finite size effects which appear to be nontrivial \cite{GusyninReenders}.
Also, the running of the coupling $\alpha$, due to the running of the Fermi
velocity $v_{F}$, may change the shape of the vertical part of the critical line. These
effects  most likely change the kind of phase transition to the second order one.
This would indicate that the semimetal-insulator transition in graphene is likely
to be of second order.
We expect that the form of the critical curve in graphene can be checked in further
lattice simulations.

Also, our results maybe important for the proper interpretation
of lattice simulations of low-energy field-theory model for quasiparticles in graphene
interacting through the long-range Coulomb potential \cite{Drut} because local
four-fermion terms are expected to be generated by the lattice regularization
procedure. We showed that in spite being small ($G\sim$ lattice constant)
the induced local interactions can play a significant role in the critical
behavior observed in lattice simulations. A related aspect of the near-critical
behavior is the appearance of composite electron-hole degrees of freedom whose
form factors slowly decrease with  momentum (tight bound states), and
the momentum behavior is governed by large anomalous dimension.
Their dynamics can be studied similarly to that in strongly
coupled QED \cite{composite-degrees} but this remains a problem for future
investigations.

\section{Acknowledgments}
We are grateful to I.F.~Herbut, A.B. Kashuba, V.M. Loktev and V.A. Miransky for useful
discussions. The present work was supported partially by the SCOPES grant
No. IZ73Z0\verb|_|128026 of Swiss NSF,
by Ukrainian State Foundation for Fundamental Research under Grant No. F28.2/083,
and by the Program of Fundamental Research of the Physics and Astronomy Division of
the NAS of Ukraine.


\begin{thebibliography}{99}
\bibitem{footnote1}The excitonic instability in the many-body problem was also studied
in a recent paper \cite{Fertig}. The authors search for poles in the
particle-hole vertex function and find that the poles appear for
$\alpha>\alpha_{c}=1/2$. This method is in fact close to that used
in \cite{excitonic-instability} where homogeneous BS equation for
the bound state wave function (residue at the pole) has been solved.
Note, however, that the  polarization was not taken into account in
\cite{Fertig}, therefore their critical coupling $\alpha_{c}=1/2$
differs from that obtained in \cite{excitonic-instability} where
$\alpha_{c}=1.62$.
\bibitem{footnote3}For two-body problem, solving the Breit equation,  the authors
of Ref.\cite{Guinea}  found the critical coupling twice bigger than
that for the one-body Coulomb impurity problem. This difference of
critical couplings for the Breit and Dirac equations in the Coulomb
potential is well known in $3+1$ dimensions, compare, for example,
Refs.\cite{FominMiransky} and \cite{Bawin}.
\bibitem{footnote4}The influence of Fermi velocity renormalization
on a gap generation was studied in a recent paper
\cite{Khveshchenko1}. It was shown there that the momentum dependent
velocity  increases the critical value of the coupling constant.
\bibitem{Fock-term}To take into account the Fock contribition one should replace
$4N_{f}$ by $4N_{f}-1$ in Eq.(\ref{phasediagram}).
\end{thebibliography}
\end{document}